\documentclass[11pt]{article}
\usepackage[utf8]{inputenc}
\usepackage{enumerate}
\usepackage{xcolor}
\usepackage{amsmath}
\usepackage{amsfonts}
\usepackage{amssymb}
\usepackage{capt-of}
\usepackage[a4paper]{geometry}
\usepackage{float}
\usepackage{graphicx}
\usepackage{authblk}
\usepackage{authblk}
\usepackage{graphics, setspace}
\usepackage[title]{appendix}
\usepackage{hyperref}
\newcommand{\mathsym}[1]{{}}
\newcommand{\unicode}[1]{{}}

\newcommand\ii{\'{\i}}

\begin{document}

\title{Matching high  and low temperature regimes of massive scalar fields}

\author[1]{Manuel Asorey}
\author[1]{Fernando Ezquerro}

\affil[1]{\ Departamento de F\'isica Te\'orica, Centro de Astropart\'{\i}culas y F\'{\i}sica de Altas, Energ\'{\i}as, Universidad de Zaragoza, 50009 Zaragoza, Spain}

\date{}
\maketitle

\begin{abstract}
We analyze the matching of  high  and low temperature expansions of the effective action of massive
scalar fields confined between two infinite walls with different boundary conditions.
 One remarkable low temperature effect is the exponential decay of the vacuum energy
 with the separation of the walls and the fact that the rate of decay is half for  the boundary
 conditions which involve a connection between the boundary conditions of  the two walls.
 In particular, the rate for Dirichlet boundary conditions is double than that of periodic boundary conditions.
\end{abstract}

	\section{Introduction}
	
	In the last years there has been an increasing interest in the study of  Casimir energy
	of gauge theories on bounded domains for different boundary conditions. There are two interesting reasons for
	doing it. First, because not too much is known about the effects for self-interactions on the Casimir energy
	 (see e.g. for pioneer analysis \cite{Symanzik:1981wd, BORDAG1985192}) and
	the second reason is that in the ultraviolet regime non-abelian pure gauge theories involves
	 free fields because of asymptotic freedom and the perturbative calculations can be achieved analytically. 
	 However, in the infrared
	the theories become massive and radical  changes are then expected due the non-perturbative effects in Casimir
	energy because the confinement mechanism dramatically changes the properties of the quantum vacuum.
	
	Moreover it has been conjectured that  the low energy spectrum  of  2+1 dimensional  Yang-Mills theory is driven by a 
	massive scalar field \cite{KARABALI1998661}, which to a certain extend has been numerically confirmed \cite{PhysRevLett.121.191601,PhysRevD.98.105009} for Dirichlet boundary conditions. If this analogy is correct
	then the comparison with the results for other type of boundary conditions will clarify this picture.
	
	In any case the numerical verification of this conjecture requires to know the dependence of the effective action
	of massive scalar field theories at finite temperatures because the numerical simulations always are developed in
	such a framework.  An analysis in low temperature regime of the theory	\cite{asorey2025new1} (see \cite{hays197932, Mera_2015} for lower dimensions cases)
	points out that the vacuum energy is exponentially damped as the distance between the walls increase. 
	But the decaying rate depends on the type of boundary conditions. 
	
	%In general, if  characterizes two types of boundary conditions
	In this paper, we analyse the temperature dependence of the effective action of massive free theories and   how well the high and low temperature regimes match each other. In particular, we scrutinize the dependence of those regimes on the boundary conditions and check that they 	experiment the same type of exponential decay rates.

	We compare the results with those obtained the low and high temperature regimes of massless theories  \cite{munoz2020thermal,Brevik_2006} which can be obtained in the limit $m\to 0$ in a continuous way.
We remark that Casimir vacuum energy for massless theories   shows a more complex dependence of the Casimir energy on the parameters of the boundary conditions \cite{Boundary_general_2013}.

\section{Boundary conditions for scalar fields}

The most general boundary conditions of a massive scalar field $\phi$ confined between two homogeneous parallel plates  separated by a distance $L$ at finite temperature $T$ in three spatial dimensions are given in the Euclidean time $\tau=it $ formalism by
%of $2\times 2$ unitary matrices $U$ \cite{asorey2005global} 
\begin{equation*}
	\phi(\tau+{2\pi}/{T}, \mathbf{x})=\phi(\tau, \mathbf{x});    \psi - i\ell\dot\psi = U(\psi +i\ell\dot\psi ),
	\label{bccc}
\end{equation*}
where 
\begin{equation*}
	\psi=\begin{pmatrix}
		\psi(L/2)  \\ \psi(-L/2)  \\ 
	\end{pmatrix},
	\dot\psi=\begin{pmatrix}
		\dot\psi(L/2)  \\ \dot\psi(-L/2)  \\ 
	\end{pmatrix}; U\in \text{U(2)},
\end{equation*}
 $\psi(\pm L/2)= \phi(\tau,x_1,x_2, \pm  L/2)$ being the values of  
the fields  $\phi$ at the boundary plates,  $\dot\psi(\pm L/2)=\pm\partial_3(\phi(\tau,x_1,x_2, \pm L/2))$  the outward normal  derivatives on the plates, and $\ell$  an arbitrary scale parameter that we shall set from now on $\ell=1$ for simplicity.

We shall use the parametrization of $U\in U(2)$ matrices given by
\begin{equation*} \label{U2}
U(\theta,\eta,{{\bf n}})={\mathrm e}^{i\theta}\left(\mathbb{I }\cos\eta+i{{\bf n}}\cdot\boldsymbol{\sigma}\,\sin\eta \right);\hspace{0.4cm}
 \theta\in[0,2\pi),\eta\in[-\pi/2,\pi/2),  {{\bf n}}\in S^2\nonumber
\end{equation*}
 where  $\boldsymbol{\sigma}$  are the Pauli matrices.  Also, we have to impose that $n_2=0$ for the case of real scalar fields.
 
 The effective action in the Euclidean time is given by
 \begin{equation} \nonumber \label{s.long}
	S_{\hbox{eff}}(\beta)=-\log Z(\beta)
\end{equation}
in terms of  the partition function
\begin{equation}\nonumber
	Z(\beta)=\hbox{det}\left(-\partial_\tau^2-\nabla^2+m^2\right)^{1/2}
\end{equation}
where $m$ is the mass of the scalar field.

The determinant can be computed in terms of the eigenvalues $\lambda$ of the differential operator $-\partial_\tau^2-\nabla^2+m^2$
\begin{equation}\nonumber
	\lambda= \left(\frac{2\pi l}{\beta}\right)^2+k^2+\kappa_i^2+m^2 ;\hspace{0.4cm} l \in \mathcal Z, i=0,1\ldots
\end{equation}
given by the temporal modes associated to the Matsubara frequencies, the continuous spatial modes $k^2=k_1^2+k_2^2 $ of the two dimensional unbounded coordinates $x_1,x_2$ and the discrete spatial modes associated to the transversal modes in $x_3$ which depend on the boundary conditions.

Now, the partition function can be written in terms of the the zeta function regularization as
\begin{equation}\nonumber
\log Z(\beta)=-\left.\frac{1}{2}\frac{d}{ds}\zeta \left(s\right)\right|_{s=0}
\end{equation}
with
\begin{equation}\nonumber
	\zeta \left(s\right)=
\frac{A \mu^{2s}}{4\pi(s-1)}\sum_{l,i} \textstyle\left(\left(\frac{2\pi l}{\beta}\right)^2\!\!\!+k_i^2+m^2\right)^{-s+1} \label{ini3d}
\end{equation}
where we have integrated the continuous spatial part, regularized the ultraviolet divergence using the analytic extension of the zeta function and introducing a renormalization scale parameter $\mu$ and $A$ is the area of the plates.  By using Cauchy formula we can
compute the sum of the spatial eigenvalues in the zeta function by the contour integral
\begin{equation}\nonumber
\sum_{l=-\infty}^\infty \oint  \left[\left(\frac{2\pi l}{\beta}\right)^2+k_i^2+m^2\right]^{-s+1}\frac{d}{dk}\log h_U(k),
\end{equation}
where the contour is reduced to a thin strip around the positive real axis and $h_U(k)$ is the spectral function
\begin{equation}\nonumber %\label{spectral_3d
	h_U(k)  =  2i {\mathrm e}^{i\theta}\left(\sin(kL)\left((k^2\!\!-\!\!1)\cos \eta \right.+(k^2\!\!+\!\!1) \cos \theta \right)
	 -2k\left.\sin \theta \cos (kL)-2kn_1\sin \eta \right)%k^{-1}
	\nonumber
\end{equation}
whose zeros are the discrete eigenvalues $\kappa_i^2$ of the operator $-\partial_3^2$ satisfying the boundary conditions
at  $x_3=\pm L/2$.

In order to find analytic formulas for the effective action we consider two
different extreme temperature regimes.
\section{High temperature regime}

After a renormalization of the ultraviolet divergences the effective action become finite
but we have a different expansion for the boundary conditions that have zero-modes and the 
ones that do not. The difference between these two cases also affect to the Heat kernel 
expansion coefficients  \cite{munoz2015qft}.

The results in the case with zero-modes reads
\begin{align}
\label{3d_highT_smal-1}
	S^0_{\mathrm{eff}}=&-\frac{ALm^3}{12\pi}-\frac{A L \pi^2 }{90\beta^3 }+\frac{AL m^2 }{24 \beta}\displaystyle{ -\frac{AL \beta m^4}{32\pi^2 }\left(\gamma+\log \frac{\mu \beta}{4\pi }\right)}\\  &-
	\frac{4\pi^2 AL}{3\beta^3}\sum_{n=3}^{\infty} \frac{\Gamma\left(\frac{5}{2}\right)}{n!\Gamma\left(\frac{5}{2}-n\right)}\left(\frac{m\beta}{2\pi }\right)^{2n}\zeta_R\left(2n-3\right)\\ 
	& -\frac{A}{8\pi }
	\int_{m}^\infty \ dk \left(m^2-k^2\right)\left(L-\frac{d}{dk}\log\frac{ h^L_U(ik)}{h^{\infty}_U(k)}\right)\label{3d_highT_smal-2}\\ 
	&-\frac{A}{4\pi }\displaystyle{\sum_{l=0}^\infty\int_{_{\sqrt{\left(\frac{2\pi l}{\beta}\right)^2+m^2}}}^\infty  dk} \left(\left(\frac{2\pi l}{\beta}\right)^2+m^2-k^2\right)
 \left(L-\frac{d}{dk}\log\frac{ h^L_U(ik)}{h^{\infty}_U(k)}\right), \label{3d_highT_smal-3}
\end{align}
where
%$\textstyle {h^{\infty}_U(k)={k}^{-1}\left((k^2+1)\cos \theta +(k^2-1)\cos \eta +2k\sin\eta\right)} $
\begin{eqnarray}& \textstyle { h^{\infty}_U(k)=(k^2+1)\cos \eta +(k^2-1)\cos \theta}+2k\sin\theta \nonumber
%\textstyle { h^{\infty}_U(k)=(k+1/k)\cos \beta +(k-1/k)\cos \alpha}\nonumber\\
 %&+2\sin\alpha \nonumber
\end{eqnarray}
and $\mu$ is the renormalization scale parameter.

The second term in the expansion (\ref{3d_highT_smal-1}) of $S^0_{\mathrm{eff}}$  is the universal Boltzmann term of the  high temperature expansion.  The rest of the $L$-linear dependent terms depend in positive powers of  the mass, meaning that they vanish in the massless theory. The fourth term (\ref{3d_highT_smal-1}) has a logarithm dependence on  the free   renormalization scale parameter $\mu$. This is the only ambiguity of the calculation that has to fixed by the renormalization conditions. Notice that this term  is proportional to the space-time volume ($AL\beta$) giving a physical meaning to its renormalization as a vacuum energy. The first term that is defined by an spectral integral (\ref{3d_highT_smal-2}) corresponds to the Casimir energy in a field theory in 2+1 dimensional space-time with the same boundary conditions. The $L$ dependence of the last two terms  (\ref{3d_highT_smal-2}) (\ref{3d_highT_smal-3}) is not linear. In fact, both  are exponentially suppressed as $e^{-mL}$ in the large $L$ limit. The appearance of renormalization group ambiguities encoded in the $\log \mu$ contributions are intrinsically associated 
to the existence of zero modes \cite{asorey_temp1}.

In the case without zero-modes the result has some extra terms
\begin{equation}\label{3d_highT_small_no_zero}\nonumber
	S_{\mathrm{eff}}= S^0_{\mathrm{eff}}+\frac{Am^2}{8\pi}\left(\log (m\beta)-\frac{1}{2}\right) +\frac{A\zeta_R(3)}{4\pi\beta^2} +\frac{A\pi}{\beta^2}\sum_{n=2}^\infty\frac{(-1)^{n+1}\zeta_R(2n-2)}{n(n-1)}\left(\frac{m\beta}{2\pi}\right)^{2n}.
	\end{equation}

If we take the limit $m=0$ the effective action reduces to:
\begin{align}\nonumber
	S^0_{\mathrm{eff}} = &  -\frac{A L \pi^2 }{90\beta^3 }-\frac{A}{8\pi }
	\int_{m}^\infty  dk\ k^2\left(L-\frac{d}{dk}\log\frac{ h^L_U(ik)}{h^{\infty}_U(k)}\right)\\ \nonumber
	&+\frac{A}{4\pi }\sum_{l=0}^\infty\int_{\frac{2\pi l}{\beta}}^\infty \ dk \left(\left(\frac{2\pi l}{\beta}\right)^2-k^2\right) 
	\left(L-\frac{d}{dk}\log\frac{ h^L_U(ik)}{h^{\infty}_U(k)}\right),
\end{align}
in the case with zero-modes,
and 
\begin{eqnarray}\nonumber
	S_{\hbox{eff}}= S^0_{\mathrm{eff}} +\frac{A\zeta_R(3)}{2\pi\beta^2}
\end{eqnarray}	
for the no zero-modes case.

\section{Low temperature regime}

In this case we exploit the use Schwinger integral representation and Poisson formula for resumming the Matsubara temporal  modes. The result is

\begin{align}\label{Spectra_lowT_3d_zero}\nonumber
	&S^0_{\mathrm{eff}}=\frac{A\beta }{12\pi^2  }\int_{m}^\infty  dk (k^2-m^2)^{3/2}\left(L-\frac{d}{dk}\log \frac{h^L_U(ik)}{h^{\infty}_U(k)}\right) -\frac{AL\beta m^4}{32\pi^2}\left(\log \frac{\mu}{m}+\frac{3}{4}\right)\\ \nonumber
	& -\frac{A}{2\pi \beta^2}\left(\beta m \hbox{Li}_2\left(e^{-m\beta}\right)+\hbox{Li}_3\left(e^{-m\beta}\right)\right)\\\nonumber
	&-\!\!\frac{i A}{4\pi^2  \beta^2}\!\!\int_{0}^\infty\!\!\!\!  dr %\\ \nonumber
	\!\!\left[\!\left( \beta\sqrt{r^2 e^{2i\sigma}\!\!+\!m ^2}\,  \mathrm{Li}_2(e^{-\sqrt{r^2e^{2i\sigma}+m^2}\beta })\!	 + \!\textstyle{\mathrm{Li}_3(e^{-\sqrt{r^2e^{2i\sigma}+m^2}\beta })}\right)\!\frac{d}{dr}\!\log (\frac{h_U(re^{i\sigma})}{r^3})\right.\\
\nonumber	& %\frac{A}{4\pi^2 i \beta^2}\int_{0}^\infty dr 
	-\!\!\left.\left( \beta\sqrt{r^2 e^{-2i\sigma}+m ^2}\  \hbox{Li}_2(e^{-\sqrt{r^2e^{-2i\sigma}+m^2}\beta })\!+ \hbox{Li}_3(e^{-\sqrt{r^2e^{-2i\sigma}+m^2}\beta })\right)\!\frac{d}{dr}\!\log (\frac{h_U(re^{-i\sigma})}{r^3})\right].
\end{align}
for boundary conditions with zero-modes. From the first term of  this expression  one can get the Casimir energy.
The second term is similar to the one obtained in the high temperature regime that depends on the spacetime volume ($AL\beta$) and the renormalization free parameter $\mu$. Finally the remaining  terms  
 are exponentially suppressed as $e^{-m\beta}$ in the low temperature regime $T=1/\beta$ limit.
 
In the case without zero-modes the result has some extra terms
\begin{align}%\label{Spectral_lowT_3d}
	\nonumber
	&S_{_{\hbox{eff}}}= \frac{A\beta }{12\pi^2  }\int_{m}^\infty   dk (k^2\!-\!m^2)^{3/2}\left[L\!-\!\frac{d}{dk}\log \frac{h^L_U(ik)}{h^{\infty}_U(k)}\right] -\frac{AL\beta m^4}{32\pi^2}\left(\log \frac{\mu}{m}+\frac{3}{4}\right)+\frac{A\beta m^3}{24}\nonumber\\\nonumber
&-\!\!\frac{i A}{4\pi^2  \beta^2}\!\!\int_{0}^\infty\!\!\!\!  dr %\\ \nonumber
	\!\!\left[\!\left( \beta\sqrt{r^2 e^{2i\sigma}\!\!+\!m ^2}\,  \mathrm{Li}_2(e^{-\sqrt{r^2e^{2i\sigma}+m^2}\beta })\!	 + \!\textstyle{\mathrm{Li}_3(e^{-\sqrt{r^2e^{2i\sigma}+m^2}\beta })}\right)\!\frac{d}{dr}\!\log (\frac{h_U(re^{i\sigma})}{r})\right.\\
\nonumber	& %\frac{A}{4\pi^2 i \beta^2}\int_{0}^\infty dr 
	-\!\!\left.\left( \beta\sqrt{r^2 e^{-2i\sigma}+m ^2}\  \hbox{Li}_2(e^{-\sqrt{r^2e^{-2i\sigma}+m^2}\beta })\!+ \hbox{Li}_3(e^{-\sqrt{r^2e^{-2i\sigma}+m^2}\beta })\right)\!\frac{d}{dr}\!\log (\frac{h_U(re^{-i\sigma})}{r})\right].
\end{align}
	%&+\frac{A}{4\pi^2 i \beta^2}\int_{0}^\infty dr \left( \beta\sqrt{r^2 e^{2i\sigma}+m ^2}\  \hbox{Li}_2\left(e^{-\sqrt{r^2e^{2i\sigma}+m^2}\beta }\right)+\hbox{Li}_3\left(e^{-\sqrt{r^2e^{2i\sigma}+m^2}\beta }\right)\right)\frac{d}{dr}\log (f_U(re^{i\sigma}))\\
%%%&\left[\!\left( \beta\sqrt{r^2 e^{2i\sigma}\!\!+\!m ^2}\,  \mathrm{Li}_2(e^{-\sqrt{r^2e^{2i\sigma}+m^2}\beta }) + \mathrm{Li}_3\left(e^{-\sqrt{r^2e^{2i\sigma}+m^2}\beta }\right)\right)\frac{d}{dr}\log (\frac{h_U(re^{i\sigma})}{r })\right.\\
%%%\nonumber	& %\frac{A}{4\pi^2 i \beta^2}\int_{0}^\infty dr 
%%%	-\!\left.\left( \beta\sqrt{r^2 e^{-2i\sigma}+m ^2}\  \hbox{Li}_2\left(e^{-\sqrt{r^2e^{-2i\sigma}+m^2}\beta }\right)  + \hbox{Li}_3\left(e^{-\sqrt{r^2e^{-2i\sigma}+m^2}\beta }\right)\right)\frac{d}{dr}\log (\frac{h_U(re^{-i\sigma})}{r })\right]
with a slightly different spectral functions terms.  Let us now finish the explicit calculations for two very different types of
boundary conditions: Dirichlet and periodic.

\section{Dirichlet Boundary conditions}

In this case the spectrum of the Laplacian operator is  $k_1^2 +k_2^2+\kappa_j^2$ with $j=1,2,\cdots $ and
$\kappa_j=\pi j/L$. There  are not zero-modes with this boundary conditions and 
\begin{eqnarray}\nonumber %\label{spectral_3d
	h_d(k)=  4i \sin(kL).\nonumber
\end{eqnarray}
In this case the integrals over the spectral function in the expression of the effective action expansion in high temperature  can be analytically computed and the final result is
\begin{align}\nonumber
	&S_{\mathrm{eff}} =\!\frac{A L m^2 }{24\beta}-\frac{Am^3L}{12\pi}-\frac{A L \pi^2 }{90\beta^3 }-\frac{A L\beta m^4}{32\pi^2 }(\gamma\!+\!\log \frac{\mu \beta}{4\pi })\\\nonumber
	&-\frac{4\pi^2AL}{3\beta^3}\sum_{n=3}^{\infty}\frac{\Gamma\left(\frac{5}{2}\right)}{n!\Gamma\left(\frac{5}{2}-n\right)}\left(\frac{m\beta}{2\pi }\right)^{2n}\zeta_R\left(2n-3\right) +\frac{A\zeta_R(3)}{4\pi\beta^2}+\frac{Am^2}{8\pi}\left(\log (m \beta)-\frac{1}{2}\right)\\  \nonumber
	&+\frac{A\pi }{2\beta^2}\sum_{n=2}^\infty\frac{(-1)^{n}\zeta_R(2n-2)}{n(n-1)}\left(\frac{m\beta}{2\pi }\right)^{2n} -\frac{A}{16L^2\pi}\left(2mL\ \hbox{Li}_2(e^{-2Lm})+\hbox{Li}_3(e^{-2Lm})\right)\\ \nonumber
	&-\frac{A}{8\pi L^2}\sum_{l=0}^{\infty}\left[ \hbox{Li}_3(e^{-\frac{2L}{\beta} \sqrt{(m\beta)^2+(2\pi l)^2}})+\frac{2L}{\beta}\sqrt{\left(m\beta \right)^2+\left(2\pi l \right)^2}  \hbox{Li}_2\left(e^{-\frac{2L}{\beta} \sqrt{\left(m\beta \right)^2+\left(2\pi l \right)^2} }\right)\right].\nonumber
\end{align}
In the low temperature regime the result is
\begin{align}\nonumber
	S_{_{\hbox{eff}}}=&-\frac{A\beta L m^4}{32\pi^2}\left(\log \frac{\mu }{m}+\frac{3}{4}\right)-\frac{A \beta m^2 }{8\pi^2 L }\sum_{j=1}^\infty  \frac{K_{2}( 2mL j)}{j^2}+\frac{A\beta m^3}{24\pi}\\ \nonumber
	 -&\frac{A}{2\pi \beta^2}\sum_{j=1}^\infty\left[ \hbox{Li}_3\left(e^{-\sqrt{(\frac{\pi j}{L})^2+m^2}\beta}\right) +\beta\sqrt{\left(\frac{\pi j}{L}\right)^2+m ^2}\  \hbox{Li}_2\left(e^{-\sqrt{(\frac{\pi j}{L})^2+m^2}\beta }\right) \right] \nonumber
\end{align}
which shows from the leading terms of the series  of modified Bessel functions $K_{2}\left( 2 mL j\right)$  induce an exponential decay $E_c\approx- \mathrm{e}^{-2mL}$ of the Casimir energy.

In the regime $mL<2\pi$ this expression can be rewritten as
\begin{align}\nonumber
	S_{_{\hbox{eff}}}&= -\frac{A \beta\ \pi^2}{1440 L^3 }+\frac{A \beta m^2  }{96 L }-\frac{A \beta\ m^4L }{32\pi^2 }\left(\gamma +\log{\frac{\mu L}{2\pi }}\right)\\\nonumber
	&-\frac{A \beta\ \pi^2}{12 L^3}\sum_{n=3}^\infty \frac{\Gamma(\frac{5}{2})}{n!\Gamma(\frac{5}{2}-n)}\left(\frac{mL}{2\pi}\right)^{2n}\zeta_R\left(2n-3\right)\\
	&  %+\frac{A\beta m^3}{24\pi}
	 -\frac{A}{2\pi \beta^2}\sum_{j=1}^\infty\left[  \hbox{Li}_3\left(e^{-\sqrt{(\frac{\pi j}{L})^2+m^2}\beta}\right)+\beta\sqrt{\left(\frac{\pi j}{L}\right)^2+m ^2}\  \hbox{Li}_2\left(e^{-\sqrt{\left(\frac{\pi j}{L}\right)^2+m^2}\beta }\right) \right] \nonumber
\end{align}
which  in the limit of $m=0$ and $\beta \rightarrow \infty$ gives rise the well known value for the Casimir Energy  $E_c={\pi^2}/{1440 L^3}$ of a massless scalar free field theory \cite{Bordag_2001, milton2001casimir}.

\begin{figure}[H]
	\centering
	\includegraphics[width=0.8\textwidth]{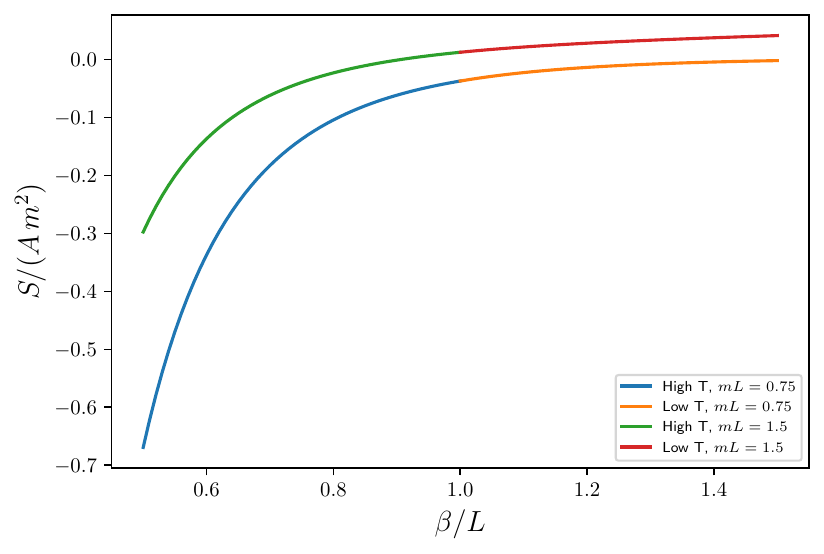}
	\caption{Perfect matching between the low and high temperature regimes for the effective action with Dirichlet boundary conditions}\label{dirichlet_3d}
\end{figure}

It is also remarkable how well the two behaviors a high and low temperature match at intermediate values of the temperature (see fig. \ref{dirichlet_3d}) .  

%The dependence of this effective action with respect to the temperature follows the

\section{Periodic Boundary conditions}
In this case the spectrum of the Laplacian operator is  $k_1^2 +k_2^2+\kappa_j^2$ with $j=0, \pm 1,\pm 2,\cdots $ and
$\kappa_j=2\pi j/L$.
With this boundary conditions there are zero-modes ($j=0$) and  the spectral function is
\begin{equation*} %\label{spectral_3d
	h_p(k)=  4k( \cos kL-1).
\end{equation*}

In this case also the integrals over the spectral function in the expression of the effective action expansion in high temperature  can be analytically computed giving the result
\begin{align}\nonumber
S_{\mathrm{eff}}& =\frac{A L m^2 }{24\beta}-\frac{Am^3L}{12\pi}-\frac{A L \pi^2 }{90\beta^3 }-\frac{A L\beta m^4}{32\pi^2 }(\gamma+\log \frac{\mu \beta}{4\pi })\\\nonumber
	&-\!\frac{4\pi^2AL}{3\beta^3}\sum_{n=3}^{\infty}\frac{\Gamma\left(\frac{5}{2}\right)}{n!\Gamma\left(\frac{5}{2}-n\right)}\left(\frac{m\beta}{2\pi }\right)^{2n}\!\!\zeta_R\left(2n-3\right)
	%&+\frac{A\zeta_R(3)}{4\pi\beta^2}+\frac{Am^2}{8\pi}\left(\log (m \beta)-\frac{1}{2}\right)+\frac{A\pi }{2\beta^2}\sum_{n=2}^\infty\frac{(-1)^{n}\zeta_R(2n-2)}{n(n-1)}\left(\frac{m\beta}{2\pi }\right)^{2n}\\\nonumber
	-\!\frac{A}{2L^2\pi}\left(mL\ \hbox{Li}_2(e^{-Lm})+\hbox{Li}_3(e^{-Lm})\right)\\
	&\nonumber-\frac{A}{\pi L^2}\sum_{l=1}^{\infty}\left[ \hbox{Li}_3\left(e^{-L \sqrt{m^2+(\frac{2\pi l }{\beta})^2}}\right) 
% \!\!\!\!\!\!\!\!\!\!\!\!\!\!\!\!\!\!\!\!+\left.\frac{2L}{\beta}\textstyle{ \sqrt{\left(m\beta \right)^2+\left(2\pi l \right)^2}}  \hbox{Li}_2(e^{-\frac{2L}{\beta} \sqrt{\left(m\beta \right)^2+(\frac{2\pi l }{\beta})^2} })\right].\nonumber
%	&-\frac{A}{16L^2\pi}\left(2mL\ \hbox{Li}_2\left(e^{-2Lm}\right)+\hbox{Li}_3\left(e^{-2Lm}\right)\right)\\
%	&-\frac{A}{8\pi L^2}\sum_{l=0}^{\infty}\left( \frac{2L}{\beta} \sqrt{\left(m\beta \right)^2+\left(2\pi l \right)^2}\  \hbox{Li}_2\left(e^{-\frac{2L}{\beta} \sqrt{\left(m\beta \right)^2+\left(2\pi l \right)^2} }\right)+\hbox{Li}_3\left(e^{-\frac{2L}{\beta} \sqrt{\left(m\beta \right)^2+\left(2\pi l \right)^2}}\right)\right)
+L\sqrt{m^2+\left(\frac{2\pi l}{\beta} \right)^2}\  \hbox{Li}_2\left(e^{-L \sqrt{m^2+(\frac{2\pi l}{\beta} )^2} }\right)\right] \nonumber,
\end{align}
which in the $m\beta>2 \pi$ can be rewritten as
\begin{align}\nonumber
	S_{_{\mathrm{eff}}}& =-\frac{A\beta L m^4}{32\pi^2}\left(\log \frac{\mu }{m}+\frac{3}{4}\right)-\frac{A L m^2 }{2\pi^2 \beta }\sum_{j=1}^\infty \frac{K_{2}\left( m\beta j\right)}{j^2}
	\\ \nonumber& 
	-\frac{A}{2L^2\pi}\left(mL\ \hbox{Li}_2\left(e^{-Lm}\right)+\hbox{Li}_3\left(e^{-Lm}\right)\right)\\
	&\nonumber-\frac{A}{\pi L^2}\sum_{l=1}^{\infty}\left[ \hbox{Li}_3\left(e^{-\frac{L}{\beta} \sqrt{(m\beta)^2+(2\pi l)^2}}\right)
% \!\!\!\!\!\!\!\!\!\!\!\!\!\!\!\!\!\!\!\!+\left.\frac{2L}{\beta}\textstyle{ \sqrt{\left(m\beta \right)^2+\left(2\pi l \right)^2}}  \hbox{Li}_2(e^{-\frac{2L}{\beta} \sqrt{\left(m\beta \right)^2+\left(2\pi l \right)^2} })\right].\nonumber
%	&-\frac{A}{16L^2\pi}\left(2mL\ \hbox{Li}_2\left(e^{-2Lm}\right)+\hbox{Li}_3\left(e^{-2Lm}\right)\right)\\
%	&-\frac{A}{8\pi L^2}\sum_{l=0}^{\infty}\left( \frac{2L}{\beta} \sqrt{\left(m\beta \right)^2+\left(2\pi l \right)^2}\  \hbox{Li}_2\left(e^{-\frac{2L}{\beta} \sqrt{\left(m\beta \right)^2+\left(2\pi l \right)^2} }\right)+\hbox{Li}_3\left(e^{-\frac{2L}{\beta} \sqrt{\left(m\beta \right)^2+\left(2\pi l \right)^2}}\right)\right)
+\frac{L}{\beta}\sqrt{\left(m\beta \right)^2+\left(2\pi l \right)^2}  \hbox{Li}_2\left(e^{-\frac{L}{\beta} \sqrt{\left(m\beta \right)^2+\left(2\pi l \right)^2} }\right)\right].\nonumber
\end{align}

In a similar way we can obtain an analytic expression for the low temperature regime

\begin{align}\nonumber
	S_{_{\mathrm{eff}}}&=-\frac{A\beta L m^4}{32\pi^2}\left(\log \frac{\mu }{m}+\frac{3}{4}\right)-\frac{A \beta m^2 }{2\pi^2 L }\sum_{j=1}^\infty \frac{K_{2}\left( mL j\right)}{j^2}\\
	&\nonumber-\frac{A}{2\pi \beta^2}\sum_{j=-\infty}^{\infty}\left[ \hbox{Li}_3\left(e^{-\beta \sqrt{m^2+(\frac{2\pi j}{L})^2}}\right)
% \!\!\!\!\!\!\!\!\!\!\!\!\!\!\!\!\!\!\!\!+\left.\frac{2L}{\beta}\textstyle{ \sqrt{\left(m\beta \right)^2+\left(2\pi l \right)^2}}  \hbox{Li}_2(e^{-\frac{2L}{\beta} \sqrt{\left(m\beta \right)^2+\left(2\pi l \right)^2} })\right].\nonumber
%	&-\frac{A}{16L^2\pi}\left(2mL\ \hbox{Li}_2\left(e^{-2Lm}\right)+\hbox{Li}_3\left(e^{-2Lm}\right)\right)\\
%	&-\frac{A}{8\pi L^2}\sum_{l=0}^{\infty}\left( \frac{2L}{\beta} \sqrt{\left(m\beta \right)^2+\left(2\pi l \right)^2}\  \hbox{Li}_2\left(e^{- \beta\sqrt{m^2+\left(2\pi l \right)^2} }\right)+\hbox{Li}_3\left(e^{- \beta\sqrt{m^2+(2\pi l)^2}}\right)\right)
+\beta \sqrt{ m^2+\left(\frac{2\pi j}{L} \right)^2}\  \hbox{Li}_2\left(e^{- \beta\sqrt{m^2+(\frac{2\pi j}{L} )^2} }\right)\right],\nonumber
\end{align}
which in the $mL<2\pi$ can be rewritten as
\begin{align}\nonumber
	S_{_{\mathrm{eff}}}&=\frac{A \beta m^2 }{24 L}-\frac{Am^3\beta}{12\pi}-\frac{A \beta \pi^2 }{90 L^3 }-\frac{A \beta L m^4}{32\pi^2 }(\gamma\!+\!\log \frac{\mu L}{4\pi })\\\nonumber
	&-\frac{4\pi^2A\beta}{3L^3}\sum_{n=3}^{\infty}\frac{\Gamma\left(\frac{5}{2}\right)}{n!\Gamma\left(\frac{5}{2}-n\right)}\left(\frac{mL}{2\pi }\right)^{2n}\zeta_R\left(2n-3\right)
%	\\
%	&-\frac{A\beta L m^4}{32\pi^2}\left(\log \frac{\mu L}{4\pi}+\gamma\right)-\frac{A \beta m^2 }{2\pi^2 L }\displaystyle{\sum_{j=0}^\infty} \textstyle{\frac{K_{2}\left( mL j\right)}{j^2}}
%	\\ \nonumber& 
%	-\frac{A}{2L^2\pi}\left(mL\ \hbox{Li}_2\left(e^{-Lm}\right)+\hbox{Li}_3\left(e^{-Lm}\right)\right)
\\\nonumber
	&-\frac{A}{2\pi \beta^2}\sum_{j=-\infty}^{\infty}\left[ \hbox{Li}_3\left(e^{-{\beta} \sqrt{m^2+(\frac{2\pi j}{L})^2}}\right)+
% \!\!\!\!\!\!\!\!\!\!\!\!\!\!\!\!\!\!\!\!+\left.\frac{2L}{\beta}\textstyle{ \sqrt{\left(m\beta \right)^2+\left(2\pi l \right)^2}}  \hbox{Li}_2(e^{-\frac{2L}{\beta} \sqrt{\left(m\beta \right)^2+\left(2\pi l \right)^2} })\right].\nonumber
%	&-\frac{A}{16L^2\pi}\left(2mL\ \hbox{Li}_2\left(e^{-2Lm}\right)+\hbox{Li}_3\left(e^{-2Lm}\right)\right)\\
%	&-\frac{A}{8\pi L^2}\sum_{l=0}^{\infty}\left( \frac{2L}{\beta} \sqrt{\left(m\beta \right)^2+\left(2\pi l \right)^2}\  \hbox{Li}_2\left(e^{-\frac{2L}{\beta} \sqrt{\left(m\beta \right)^2+\left(2\pi l \right)^2} }\right)+\hbox{Li}_3\left(e^{-\frac{2L}{\beta} \sqrt{\left(m\beta \right)^2+\left(2\pi l \right)^2}}\right)\right)
 {\beta} \sqrt{m^2+\left(\frac{2\pi j}{L}\right)^2} \   \hbox{Li}_2\left(e^{-{\beta} \sqrt{m^2+(\frac{2\pi j}{L})^2} }\right)\right],
\end{align}

The  leading terms of the series  of modified Bessel functions  $K_{2}\left( mL j\right)$ induce an exponential decay of the Casimir energy with the distance between the plates $L$  $E_c\approx- \mathrm{e}^{-mL}$ than in the Dirichlet case.

In the limit of  $m=0$ y $\beta \rightarrow \infty$ the second summand gives the known value of periodic boundary conditions for a massless scalar field $E_c={\pi^2}/{90 L^3}$.

\begin{figure}[H]
	\centering
	\includegraphics[width=0.8\textwidth]{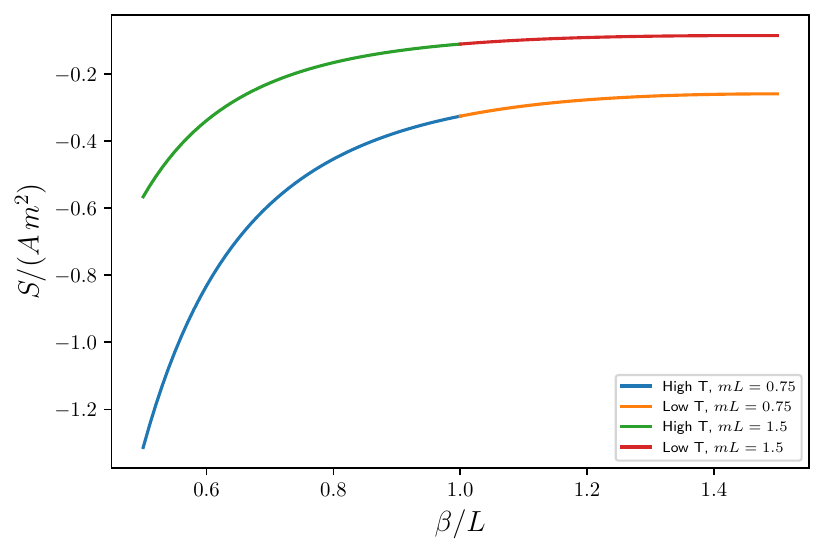}
	\caption{Perfect matching between the low and high temperature regimes for the effective action with periodic boundary conditions}\label{periodic_3d}
\end{figure}

It is also remarkable how well the two behaviors a high and low temperature match at intermediate values of the temperature as in the Dirichlet case (see fig. \ref{periodic_3d}) . 

\begin{figure}[h]
	\centering
	\includegraphics[width=0.8\textwidth]{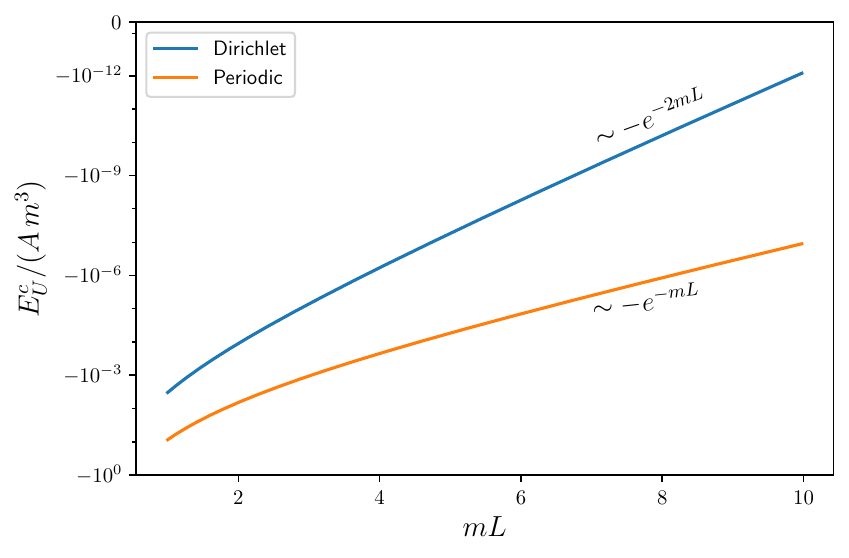}
	\caption{Exponential decay of the Casimir Energy density $E^C_U/(Am^3)$ for Dirichlet and periodic boundary conditions}\label{Casimir_3d}
\end{figure}

The novel property is that in the massive case the Casimir energy exponentially decays doubly faster for Dirichlet boundary conditions than for periodic boundary conditions (see Fig. \ref{Casimir_3d} for a comparison).

\begin{figure}[h]
	\centering
	\includegraphics[width=0.8\textwidth]{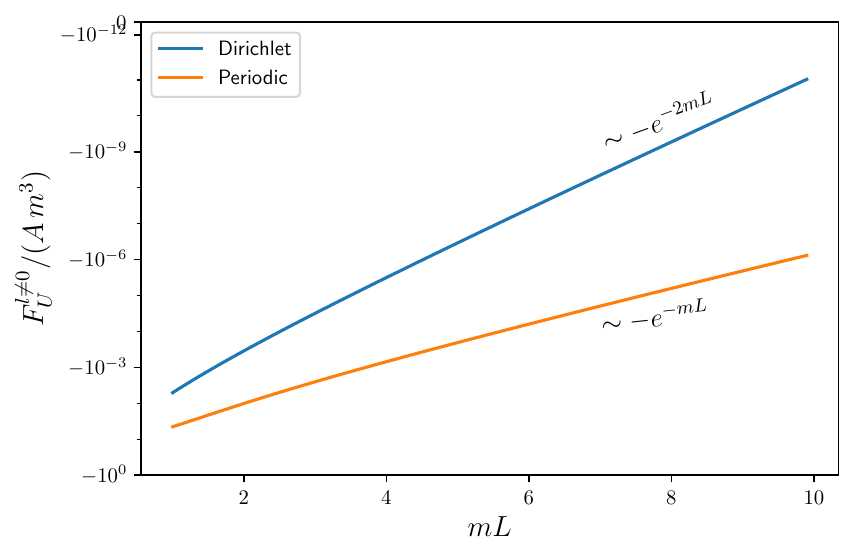}
	\caption{Exponential decay of the Free Energy density $F_U^{l\neq 0}/(Am^3)$ for Dirichlet and periodic boundary conditions, where we have subtracted in the case of Dirichlet  the extra contribution due to the absence of spectral zero-modes. Otherwise this  extra term linear dependent on L would be the leading term and  the exponential decay would not tend to zero but to the value given by this extra term. }\label {FreeEnergy_3d}
\end{figure}

The same difference on the decays holds for the rest of terms of the free energy $F=S_{_{eff}}/\beta$  at finite temperature as shown in Fig. \ref{FreeEnergy_3d}. 

\section{Conclusions}
The property that Casimir energy of massive scalar theories exponentially decay doubly faster for Dirichlet boundary conditions than for periodic boundary conditions can be extended for any type of boundary conditions. Indeed, one can show that the Casimir energy for any boundary condition  satisfying that $\mathrm{tr} \sigma_1 U=0$ decays doubly faster than for any other satisfying that $\mathrm{tr} \sigma_1 U\neq 0$ \cite{asorey2025new1}. The physical interpretation of
this feature is that the decay is faster for boundary conditions which consider the boundary walls as independent bodies, whereas for boundary conditions which interconnect the values of the fields at the two boundaries the decay rate is lower.

It is also remarkable the perfect matching of the two asymptotic regimes of high and low temperature of the 
effective action where we do have explicit analytic calculations. This is to some extend a consequence of the dual analicity
properties of the two expansions.

Now, the relevance of those results resides on the expectation that a similar behavior should emerge in non-abelian gauge theories. A  straightforward generalization of the Nair-Karabali conjecture to $3+1$ dimensions would imply a different exponential decay of the Casimir energy for different boundary conditions \cite{PhysRevD.108.014515}. If this does not agrees with   the results of this paper the interpretation becomes more involved. It might occur for instance that the scalar theory associated to the infrared sector of Yang-Mills theory is not free and its self-interactions modify the comparison with the results of the present paper. On the other hand it is also hard to explain why the {\it effective mass} that appears in the decays of the Casimir energy is much lower than the lowest glueball mass \cite{ezquerro2025casimir}. 

In any case due to the lack of analytic understanding of the confinement mechanism the analysis   of
the behavior of the vacuum energy under the effect of boundary conditions might shed some light on the underpinning structure of the non-perturbative quantum vacuum.
%%%%%%%%%%%%%%%%%%%%%%%%%%%%%%%%%%%

\section*{Acknowledgements}
We are partially supported by Spanish Grants No. PGC2022-126078NB-C21 funded by~MCIN/AEI/10.13039/ 501100011033, ERDF A way of making EuropeGrant; the Quantum Spain project of the QUANTUM ENIA of Ministerio de Econom\ii a y Transformaci\'on Digital, the Diputaci\'on General de Arag\'on-Fondo Social Europeo (DGA-FSE) Grant No. 2020-E21-17R of the Arag\'on Government, and the European Union, NextGenerationEU Recovery and Resilience Program on Astrof\ii sica y F\ii sica de Altas Energ\ii as , CEFCA-CAPA-ITAINNOVA.

\renewcommand{\refname}{References}
%\bibliographystyle{unsrt}
%\bibliography{refe}
%\bibliographystyle{eplbib}

\end{document}